\documentclass[journal]{IEEEtran}

\IEEEoverridecommandlockouts
\usepackage{cite}
\usepackage{amsmath,amssymb,amsfonts}
\usepackage{algorithm}
\usepackage{algorithmic}

\usepackage{graphicx}
\usepackage{amsfonts}
\usepackage{booktabs}
\usepackage{lipsum}
\usepackage[table,xcdraw]{xcolor}
\usepackage{xcolor}
\usepackage[letterpaper, left=0.628in, right=0.628in, bottom=1.05in,
top=0.7in]{geometry}

\usepackage{graphicx}
	\ifCLASSOPTIONcompsoc
    	\usepackage[caption=false, font=normalsize, labelfont=sf, textfont=sf]	{subfig}
	\else
\usepackage[caption=false, font=footnotesize]{subfig}
	\fi
	
\usepackage[marginal]{footmisc}
\usepackage{textcomp}
\usepackage{multirow}

\usepackage{comment}
\usepackage{blindtext}
\def\BibTeX{{\rm B\kersn-.05em{\sc i\kern-.025em b}\kern-.08em
  T\kern-.1667em\lower.7ex\hbox{E}\kern-.125emX}}

\ifCLASSINFOpdf
\else
\fi
 
\begin{document}
\title{TULVCAN: Terahertz Ultra-broadband Learning Vehicular Channel-Aware Networking}

\author{
\IEEEauthorblockN{\textbf{Chia-Hung Lin}\IEEEauthorrefmark{1}, \textbf{Shih-Chun Lin}\IEEEauthorrefmark{1}, and \textbf{Erik Blasch}\IEEEauthorrefmark{2}}\\
\vspace{1em}
\IEEEauthorblockA{\IEEEauthorrefmark{1}Intelligent Wireless Networking Laboratory, Department of Electrical and Computer Engineering,\\ 
North Carolina State University, Raleigh, NC 27695\\}
\IEEEauthorblockA{\IEEEauthorrefmark{2}Air Force Office of Scientific Research, Arlington, VA 22203\\
Email: clin25@ncsu.edu; slin23@ncsu.edu; erik.blasch.1@us.af.mil} \vspace{-0.35 in}
}
\maketitle

\begin{abstract}
Due to spectrum scarcity and increasing wireless capacity demands, terahertz (THz) communications at 0.1-10THz and the corresponding spectrum characterization have emerged to meet diverse service requirements in future 5G and 6G wireless systems. However, conventional compressed sensing techniques to reconstruct the original wideband spectrum with under-sampled measurements become inefficient as local spectral correlation is deliberately omitted. Recent works extend communication methods with deep learning-based algorithms but lack strong ties to THz channel properties. This paper introduces novel THz channel-aware spectrum learning solutions that fully disclose the uniqueness of THz channels when performing such ultra-broadband sensing in vehicular environments. Specifically, a joint design of spectrum compression and reconstruction is proposed through a structured sensing matrix and two-phase reconstruction based on high spreading loss and molecular absorption at THz frequencies. An end-to-end learning framework, namely compression and reconstruction network (CRNet), is further developed with the mean-square-error loss function to improve sensing accuracy while significantly reducing computational complexity. Numerical results show that the CRNet solutions outperform the latest generative adversarial network (GAN) realization with a much higher cosine and structure similarity measures, smaller learning errors, and  $56\backslash\%$ less required training overheads. This THz Ultra-broadband Learning Vehicular Channel-Aware Networking (TULVCAN) work successfully achieves effective THz spectrum learning and hence allows frequency-agile access. 
\vspace{1.5ex}
\end{abstract}

\IEEEpeerreviewmaketitle

\section{Introduction}
Next-generation communication systems, such as 5G and 6G, are expected to fulfill various requirements in different application scenarios in terms of data rate, latency, and power consumption, among other factors \cite{5g}. 
As a result, next-generation communication systems will likely have a requirement of 100+ Gigabits per second (Gbps) data rate requirement.
Thanks to the efforts of researchers \cite{Our1, BF}, 
a state-of-the-art communications system can already offer up to several Gbps of data rate \cite{mmWave2}. However, the Gbps rate is still an order of magnitude below that which is needed to serve popular streaming use-cases. To satisfy the urgent need of multimedia applications, there are two obvious ways of further improving the data rate: increasing the bandwidth used for communication, and improving spectral efficiency at which frequency bands are used \cite{CR, Lin4}.
In light of data rate improvements, researchers have started to focus on the utilization of terahertz (THz)-bands for next-generation communications \cite{THz1}.
Although abundant bandwidth is offered, 
due to the special properties of THz channel such as \textit{ultra-wide bands} and \textit{distance-dependent path loss}, existing physical layer algorithms must be redesigned or enhanced to fully account for the unique properties of THz spectrum communications \cite{THz1}.

Meanwhile, several research papers \cite{Lin1, Lin2} reveal that the spectrum under-utilization problem, which is caused by the existence of the idle channels, occurs in current communication systems, reducing the overall spectral efficiency.
To address this issue, spectrum management \cite{Lin1, Lin2} has been developed by detecting idle spectrum and temporarily assigning that spectrum to the demanding user, thus improving the overall spectral efficiency.
Among all the research topics in spectrum management, spectrum sensing is the most widely discussed in literature \cite{Lin1, Lin2, CSSS2, SS4} since it is a prerequisite for mitigating error propagation.
Recently, a vehicle-to-everything (V2X) data-coordination scenario \cite{Lin3} is a practical usage to employ spectrum sensing algorithms in THz communications. Typically, vehicle-to-infrastructure (V2I) connections will be assigned specific bands for the high bandwidth entertainment applications transmission (e.g., video streaming). On the other hand, vehicle-to-vehicle (V2V) connections may wish to occasionally perform safety message (e.g., vehicle position, speed and heading) transmission using idle bands assigned to the V2I connections. 
By employing THz Spectrum Sensing (TSS) algorithms in V2X communications, the underutilized spectrum can be reused and consequently leads to a better overall spectrum efficiency \cite{Lin3}. 
Moreover, TSS also has its potential 
for frequency resources allocation, heterogeneous communication system, and military usage \cite{SS4,SS3}. 

Although TSS algorithms in THz communications have potential to better utilize the frequency resources, few existing works contribute to the development of TSS algorithms for THz communications.
Most existing SS methods  \cite{Lin1, CSSS2} only focus on SS in the narrow band case. While some works \cite{Lin2, Lin1} 
concentrate on the development of wide band SS algorithms, there are almost no prior works considering physical channel effect in SS algorithm design, not even saying the unique features of THz communications.
When it comes to the ultra-wide band case like THz communications, sub-Nyquist sampling \cite{Lin2, Lin1} must be introduced to avoid costly and limiting hardware requirements. 
As a result, compressed sensing (CS) algorithms have been introduced to support spectrum reconstruction from measurements of sub-Nyquist sampling \cite{CSSS2} in the past decade.
However, they also suffer from high computational complexity to reconstruct the under-sampled signal. 
Recent research suggests that learning-based compression and reconstruction outperforms traditional CS algorithms since the local correlation is considered to reconstruct the desired output. 
Although there is a impressive work \cite{DLSS1} employing 
generative adversarial network (GAN) to aid the SS algorithms design, it still employs randomly generated sensing matrix to conduct the compression, failing to perform joint optimization of the compression and reconstruction to get the best reconstruction. 
Also, the consequently heavy overhead of the training process of GAN-based SS algorithm creates an implementation challenge to be employed in real scenario.
In conclusion, the development of a practical SS algorithm, which can be employed in real ultra-wideband communications, is an unsolved problem.




In this paper, we develop a deep learning (DL)-based spectrum reconstruction algorithm, named compression and reconstruction network (CRNet), to offer an efficient SS solution for THz communications.
The contributions of this work are listed as:
\begin{itemize}
  \item [1)] 
  Inspired by CS-based sub-Nyquist SS algorithms \cite{CSSS2}, we employ DL-based algorithm to perform efficient compression and reconstruction, offering lower computational complexity and decreasing latency for THz communications by considering local correlation of the spectrum in the optimization process.       
  \item [2)]
  As an addition to existing DL-based algorithms \cite{DLSS1}, CRNet introduces joint compression and reconstruction mechanism by designing a \textit{structured sensing matrix} and a corresponding reconstruction algorithm in a end-to-end learning manner, further improving reconstruction quality. By doing so, the training overhead can be significantly decreased to achieve when superior performance in terms of reconstruction quality.
  \item [3)]
  To the best of our knowledge, no existing works consider channel effect in the development of TSS algorithm. Simulation results demonstrate that the achieved performance of existing DL-based TSS algorithm degenerates severely in THZ communication scenarios. As an alternative, the proposed algorithm can provide robust reconstruction results even in different compression rate scenarios.
\end{itemize}








\section{System Model and Problem Description}
\subsection{System Setup and Signal Model} \label{sssec:num1}

\begin{figure}
    \centering
    \includegraphics[width=0.9\linewidth]{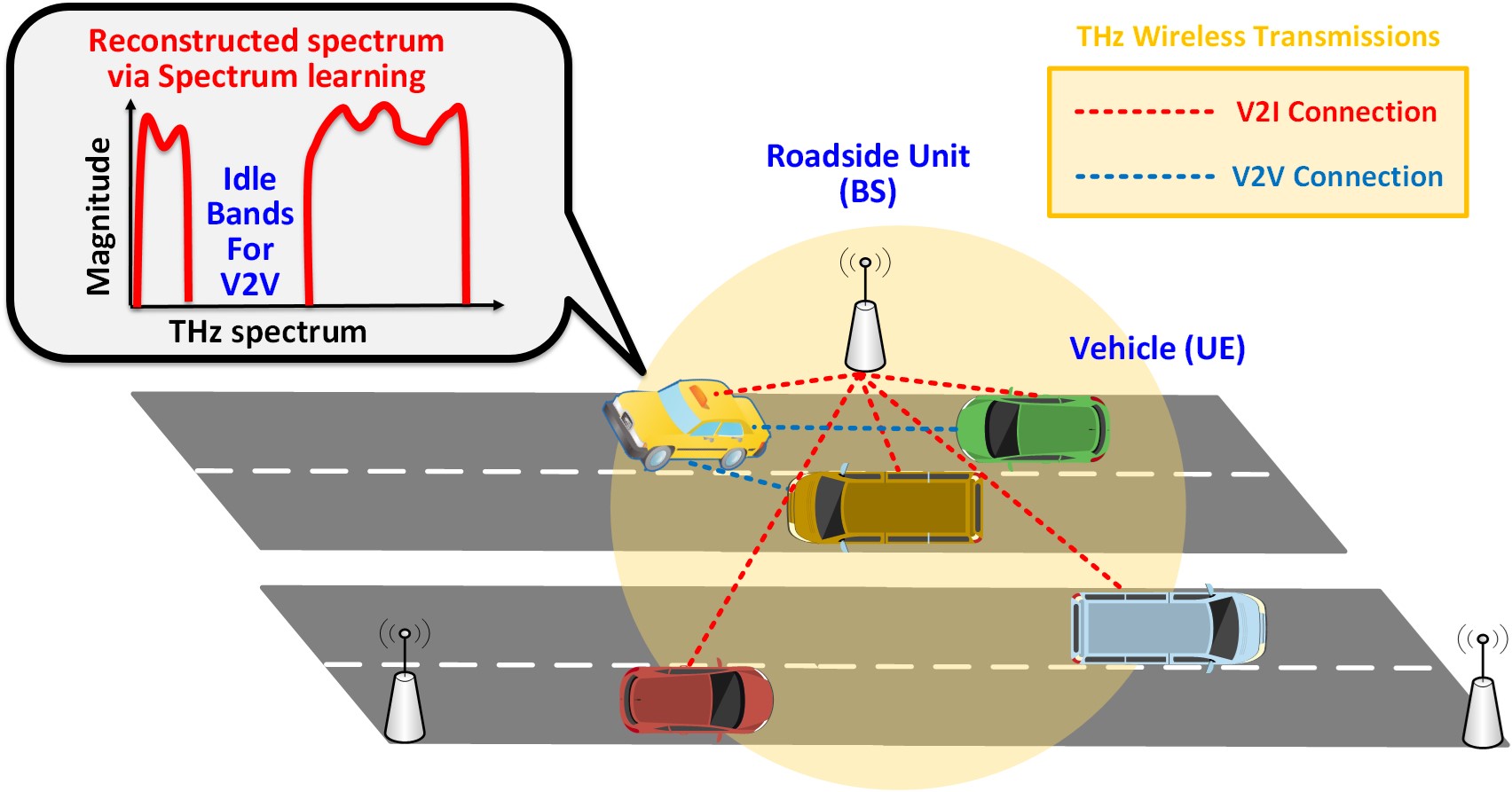}
    \caption{Considered THz communications for vehicular environment: In downlink phase, the base station (BS) will create several transmission links to different user equirements (UEs) for V2I connections. However, a UE aims to create a V2V connections with surrounding UEs to share safety messages. As a result, the UE should perform SS first to obtain the information of idle spectrum, then use those spectrum to perform V2V connections.}
    \label{fig:Scenario}
\end{figure}

As shown in Fig.~\ref{fig:Scenario}, consider a small cell including a base station (BS) with $N_r$ receiver antennas and several user equipments (UEs), each with $N_t$ transmitter antennas for vehicular communications. In the downlink phase of the considered THz-band scenario, the BS may occupy frequency bands from $f_a$ to $f_b$ by using some of the $N_s$ subcarriers to perform V2I connections. 
Considering a transmission pair between the BS and $i$ th UE, the complex baseband signal can be expressed as
\begin{equation}
\begin{aligned}
		\textbf{y}_i = \textbf{H}_i\textbf{x}_i+\textbf{n}_i,
\end{aligned}
\end{equation}
where $\textbf{y}_i \in \mathbb{C}^{N_t}$ is the received signal, $\textbf{H}_i \in \mathbb{C}^{N_t\times N_r}$ is the channel matrix, and $\textbf{x}_i \in \mathbb{C}^{N_r}$ is the transmitted signal, respectively. Assuming a perfect sampling process with a Nyquist sampling rate of $T = 1/ 2f_b$ discrete-time signals can be obtained from Eq (1).
It is assumed that the transmission between the BS-UE pair is with subcarrier $f_i$ and the distance between the BS and UE is $d_i$ in this case, thus THz channel effect $\textbf{H}_i$ can be further expressed as
\begin{equation}
\begin{aligned}
		&H_i (f_i,d_i) = H_i^{LOS}(f_i,d_i)+H_i^{NLOS}(f_i,d_i)\\
		          &= \sqrt{N_tN_r}\alpha_L(f_i,d_i)G_tG_ra_r(\theta^r_L,\phi^r_L)a_t(\theta^r_L,\phi^r_L)\\
		          &+\sqrt{N_tN_r}\sum^{n_{NL}}_{i}\alpha_i(f_i,d_i)G_tG_ra_r(\theta^r_i,\phi^r_i)a_t(\theta^r_i,\phi^r_i).
\end{aligned}
\end{equation}
In (2), $n_{NL}$ represents the number of non-line of sight (NLOS) rays. $\alpha_L(f_i, d_i)$ and $\alpha_i(f_i, d_i)$ stands for the complex channel gain of each ray. 
In each ray, $\theta / \phi$ refers to the azimuth/elevation angles of departure and arrival (AoD/AoA) and $a_t(\theta^t,\phi^t)$ and $a_r(\theta^r,\phi^r)$ are the associated array steering vectors at the transmitter and receiver sides. Finally, $G_t$ and $G_r$ is the transmit and receive antenna gains.
Furthermore, the path gain of the LOS ray can be expressed as $|\alpha_L(f_i, d_i)|^2 = L_{spread}(f_i, d_i)L_{abs}(f_i, d_i)$, where $L_{spread}(f_i, d_i)$ represents the spreading loss effect and $L_{abs}(f_i, d_i)$ represents the molecular loss effect, respectively. Given the fact that water vapor molecules cause the majority molecular absorption loss in the THz-band, $L_{abs}(f_i, d_i)$ takes the temperature, atmospheric pressure, and air density into consideration to represent the realistic THz channel behavior.
As for the path gain of the NLOS ray, we model it as $|\alpha_i(f_i, d_i)|^2 = \Gamma L_{spread}(f_i, d_i)L_{abs}(f_i, d_i)$, where $\Gamma$ is the reflection coefficient. Assuming that there are $L_1$ first-order reflected paths and $L_2$ second-order reflected paths so that $n_{NL} = L_1 + L_2$. We set the attenuated power of the first-order and second-order reflected paths as $10$dB and $20$dB, respectively.

\subsection{Problem Description}
When a  UE  aims  to  create  V2V  connections  with surrounding UEs to share safety messages, a TSS should be performed to detect the idle bands from existing V2V and V2I connections.
Yet, in a wideband scenario, to emulate the needed hardware burden, CS must be introduced to aid the reconstruction of the compressed measurements obtained from sub-Nyquist sampling.
Considering the aforementioned wideband THz communication scenario, a combining operation can be conducted at the UE to get the time domain measurements $r_i$, shown as 
\begin{equation}
\begin{aligned}
		r_i = \textbf{w}_i^{*} \textbf{y}_i = \textbf{w}_i^{*}\textbf{H}_i\textbf{x}_i+\textbf{w}_i^{*}\textbf{n}_i,
\end{aligned}
\end{equation}
where $\textbf{w}_i^{*} \in \mathbb{C}^{N_r}$ is the combining weighting. Then the time domain signals of the considered wideband system can be expressed as $\textbf{r} = \textbf{s}+\xi =  [r_1, ..., r_i, ..., r_{N_s}] \in \mathbb{C}^{N_s}$, where $\textbf{s} = [\textbf{w}_1^{*}\textbf{H}_1\textbf{x}_1,..., \textbf{w}_i^{*}\textbf{H}_i\textbf{x}_i, ...\textbf{w}_{N_s}^{*}\textbf{H}_{N_s}\textbf{x}_{N_s}]$ and $\xi = [\textbf{w}_1^{*}\textbf{n}_1,..., \textbf{w}_i^{*}\textbf{n}_i, ...\textbf{w}_{N_s}^{*}\textbf{n}_{N_s}] $.
Let $\textbf{F}$ denote a $N_s$-point discrete Fourier transform (DFT), if the signal is sampled at a sub-Nyquist rate, then the relationship between the clean spectrum $\textbf{s} \in \mathbb{C}^{N_s}$ and under-sampled measurements $\textbf{z} \in \mathbb{C}^{N_m}$ can be expressed as
\begin{equation}
\begin{aligned}
		\textbf{z} = \Phi\textbf{F}\textbf{r} = \Phi\textbf{F}(\textbf{s}+\xi),
\end{aligned}
\end{equation}
where $\Phi_{N_m \times N_s}$ is the complex-valued sensing matrix.
From Eq. (4), the goal is to design the sensing matrix and the corresponding reconstruction algorithm so that the clean spectrum $\textbf{Fs}$ can be recovered from the under-sampled measurements $\textbf{z}$ by the reconstructed spectrum $\hat{\textbf{Fs}}$. It is noteworthy that once a high quality reconstructed spectrum is available, a simple energy detector can be employed to trivially identify the unused frequency bands. Moreover, the reconstructed spectrum with high quality can enable more complex spectrum sharing design in different coexistence models of heterogeneous communication systems \cite{Lin1, Lin2}. Hence, the motivation is to develop spectrum reconstruction methods.


\section{The Development of CRNet}
We propose a compression and reconstruction network (CRNet) for efficient spectrum sensing application in THz communications. There are two features of the CRNet. 
First, conventional SS algorithms, including existing DL-based SS solutions, essentially employ randomly selected (i.e., \textit{unstructured}) sensing matrix to perform compression to get under-sampled measurements, implying there is no special design of the sensing matrix.
As an alternative, CRNet firstly introduces the joint design of compression and reconstruction by developing a \textit{structured} sensing matrix and corresponding reconstruction algorithm in a end-to-end learning manner, offering a superior performance compared to existing SS algorithms.
Secondly, compared to GAN-based SS algorithms, the training overhead of CRNet is reduced significantly. To be more specific, the under-sampled measurements obtained from the structured sensing matrix are more informative compared to that from unstructured sensing matrix, and the reconstruction can be finished by a low complexity convolutional neural network (CNN) based-model to get a promising reconstruction result.
The next section details the CRNet in terms of model architecture, loss function, and training specifics. 
\begin{figure*}
    \centering
    \includegraphics[width=0.908\linewidth]{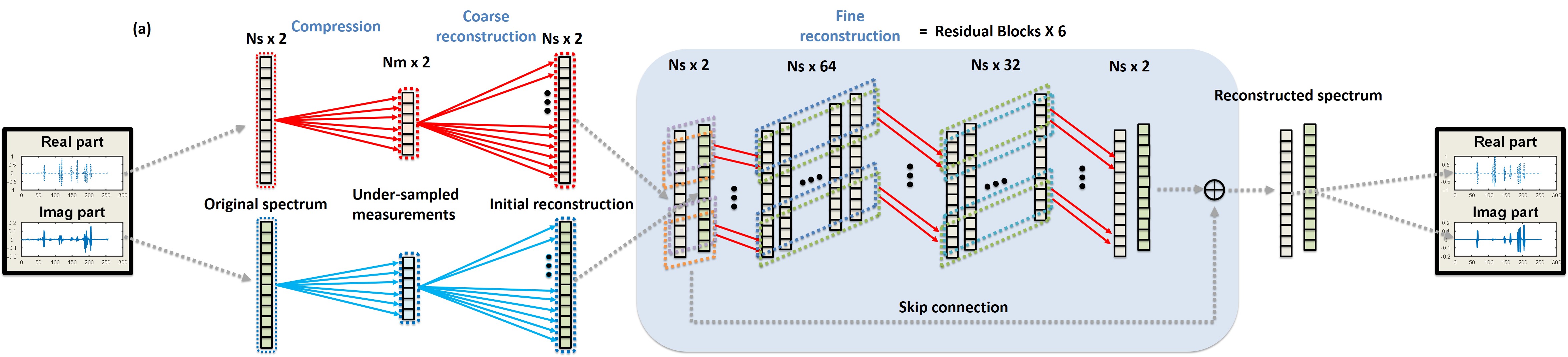}
    \caption{ The model architecture of the CRNet. In CRNet, there are three modules, compression, coarse reconstruction, and fine reconstruction. In the compression and coarse reconstruction modules, the real and imaginary part of original spectrum are compressed and reconstructed, separately. Then, the fine reconstruction module with ResNet-structure is employed to perform fine-scale reconstruction to obtain highly-fidelity reconstructed spectrum. Note that the compression and reconstruction process are performed in a end-to-end training model. As a result, the joint optimization can be performed to obtain optimal weights.}
    \label{fig:Architecture}
\end{figure*}

\subsection{Model Architecture}
As shown in Fig.~\ref{fig:Architecture}, there are three modules in CRNet, compression, coarse reconstruction, and fine reconstruction modules. 
The compression module is a specially-designed one-layer CNN to produce under-sampled measurements by making the trainable weights in the compression module act as the content of sensing matrix, being expressed as:
\begin{equation}
\begin{aligned}
		\textbf{z}_{DL} = \Phi_{DL} \textbf{Fr},
\end{aligned}
\end{equation}
where $\textbf{Fr} \in \mathbb{C}^{N_s}$ is the original spectrum, $\Phi_{DL} \in \mathbb{C}^{N_m \times N_s}$ is the sensing matrix designed by the compression module, and $\textbf{z}_{DL} \in \mathbb{C}^{N_m}$ is the under-sampled measurements from the designed sensing matrix.
Given an original spectrum \textbf{Fr}, in order to feed it into DL-based model, the input of the compression module $\textbf{Fr}$ is presented as a real vector with the size of $N_s \times 2$, containing the real part and imaginary part of the original complex vector. After the operation of the \textit{compression module}, the output $\textbf{z}_{DL}$ is a real vector with the size of $N_m \times 2$, standing for the real part and the imaginary part of the under-sampled measurements. To be more specific, a 1-dimension (1D) CNN layer is constructed with $N_m$ filters in the compression module. In each filter, the trainable weight will be created as a vector with the size of ${N_s} \times 2$, and then inner product operation between the input and the trainable weight will be conducted on the real part and imaginary part separately to obtain the computed result with the size of $1 \times 2$ representing the real part and imaginary part. This operation reflects the matrix operation between each row of the sensing matrix $\Phi_{DL}$ and the input $\textbf{Fr}$. As there are $N_m$ filters in this CNN layer, where the size of the output matches Eq. (5) to get the compressed measurements $\textbf{z}_{DL}$ for the following reconstructions.
It is noteworthy that although in the compression module, the computation is performed on the real part and imaginary part separately, the operation is exactly equivalent to the inner product on complex vector as shown in Eq. (5).
Moreover, note that there is no activation function in this 1D CNN layer to ensure the whole compression module as a linear operation.
Finally, once the training of the compression module is finished, the trainable weights in each filter (i.e., each row of the sensing matrix) can be represented as a pseudo-random (PN) sequence as shown in \cite{DLSS1}. By mixing the received signal with $N_m$ PN sequences (as there are $N_m$ rows in the sensing matrix $\Phi_{DL}$) and passing through a low-pass filter, the compressed measurements $\textbf{z}_{DL}$ can be obtained. For a real scenario, there are no implementation issues to employ the practical CRNet DL-based spectrum reconstruction algorithm.

After CRNet obtains the compressed measurements $\textbf{z}_{DL}$, the \textit{coarse reconstruction module} aims to provide an initial reconstruction for the following refinements. To do so, another 1D CNN layer is employed, which has $N_s$ filters with the size of ${N_m} \times 2$. After the CNN layer, batch normalization (BN) and parameter-Relu (PRelu) are employed to accelerate convergence and provide non-linearity, respectively. By doing so, an initial reconstruction with the size of ${N_s} \times 2$, which is the same as the original spectrum, is obtained for the following fine reconstruction module. As for the architecture of the fine reconstruction module, the ResNet-structure \cite{Res2} gradually refines the initial reconstruction result. To be more specific, the spectrum reconstruction problem is treated as a special image reconstruction problem and employs computer vision techniques to perform meticulous reconstruction.
There are two main advantages to introduce ResNet-structure into the design of the proposed spectrum reconstruction algorithm.
To explain, a typical DL model with ResNet-structure usually contains several residual blocks, which is built by several CNN layers. Instead of asking a DL model to provide a reconstruction result with high quality from scratch, the ResNet-structure lets a residual block refine the current reconstruction result based on the knowledge from previous residual blocks. As a result, all the residual blocks can coordinate with each other to synergistically produce a final reconstruction result. 
Another advantage of ResNet-structure is that a DL model with ResNet-structure is more unlikely to suffer from the over-fitting as the special skip-connection mechanism lets the DL model control the number of efficient weightings.
For the exact architecture of the fine reconstruction module, in each residual block, three 1D CNN layers with number of filters 64, 32, and 2, respectively, are built to refine the initial spectrum reconstruction. Behind each layer, BN and PRelu are also employed as the setting in the initial reconstruction module. In this paper, the CRNet comprises six residual blocks to perform fine-scale reconstruction as experiments with higher numbers of residual blocks do not improve overall performance but increase computational complexity.

From Fig. 2, a fully-CNN architecture is used in the CRNet DL-based algorithm design. There are two main reasons to use the CNN architecture. First, the needed number of trainable weights can be decreased significantly as the result of the weight sharing mechanism of the CNN. Simulation results confirm that the CRNet model outperforms existing DL models with significantly lower trainable parameters. Secondly, the CRNet fine reconstruction module aims to capture the occupied spectrum to perform fine-scale reconstruction. As the domain knowledge suggests that the occupied spectrum may appear everywhere of the whole spectrum, the convolution operation introduced by CNN can be used to captured the pattern of occupied spectrum regardless the location and number of the occupied spectrum to perform fine-scale reconstruction.

\subsection{Loss function}
An end-to-end learning is employed to jointly update all the trainable parameters in CRNet. As a result, the whole compression and reconstruction process can be designed simultaneously to achieve better performance. Let $\Theta_{CR}$ stand for the trainable weight in the coarse estimator and $\Theta_{FR}$ represent the trainable weight in the fine estimator and $f(x;\Theta_{CR}. \Theta_{FR})$ is the nonlinear transformation with $\Theta_{CR}$ and $\Theta_{FR}$, The Mean-square-error (MSE) loss function is set for the model updating, that is
\begin{equation}
\begin{aligned}
		Loss = ||\textbf{Fs}-f(\Phi_{DL} \textbf{Fr};\Theta_{CR}. \Theta_{FR})||^2.
\end{aligned}
\end{equation}
Note that during each training cycle, $\Phi_{DL}$, $\Theta_{CR}$, and $\Theta_{FR}$ will be updated jointly via the back-propagation process to gradually minimize the training loss until convergence, generating optimal structured sensing matrix and trainable weights.
Finally, as for the training specifics of the scenario in this paper, the Adam optimizer minimizes the aforementioned loss function. The initial learning rate is set as $0.0005$ and the number of epochs is set as 20. The mini-batch mechanism is employed with batch size as $128$ to facilitate fast convergence. 

\section{Simulation results}
\subsection{Data set preparation}
We follow the system model to generate spectrum samples in THz communications for DL-based algorithms training. To be more specific, we set $f_a$ and $f_b$ as $0.1$THz and $0.64$THz to employ the commonly used transmission window in THz communications. $N_s = 256$ subcarriers are equally spaced within the transmission window. Moreover, we assume that there are 8 existing users, who all chose a random, non-overlapping group of 5 subcarriers to transmit on, with at least 1 subcarriers of guard on either side. Each user ranges $1-10$ meters from the BS. 
As for the configuration of the considered communication system, $N_t = N_r =1$ and $G_t = G_r = 30$ dBi for comparable analysis in \cite{config1}. 
To implement the GAN-based algorithm, we generated the sensing matrix by randomly selecting rows from the inverse DFT matrix to generate under-sampled measurements.
As or the CRNet, the original spectrum is considered as input and the under-sampled measurements can be obtained after compression module.
Note that all the inputs and labels of the CRNet and GAN-based algorithm are normalized as [1,0] to prevent computational issues.
The number of training, validation, and testing sets are 50000, 10000, and 10000 respectively. All the results reported in this paper are the average result over the testing set.

\subsection{Performance metrics}
Three performance metrics: mean-square-error (MSE), cosine similarity, and structure similarity (SSIM), are provided to report the achieved performance of the different algorithms. 
To explain, although MSE can be used to evaluate the reconstruction quality, it cannot reflect the structure similarity between the original spectrum and the reconstructed one. As a result, two additional similarity metrics, cosine similarity and SSIM, evaluate the different algorithms in terms of reconstruction quality.
Given a reconstructed spectrum $ \hat{\textbf{Fs}} \in \mathbb{C}^{N_s} $ and the corresponding original spectrum $ \textbf{Fs} \in \mathbb{C}^{N_s} $, the MSE performance metric can be expressed as 
\begin{equation}
\begin{aligned}
		MSE = ||\textbf{Fs}-\hat{\textbf{Fs}}||^2,
\end{aligned}
\end{equation}
the cosine similarity $\rho$ can be evaluated as
\begin{equation}
\begin{aligned}
		\rho  = \frac{<\textbf{Fs},\hat{\textbf{Fs}}>}{||\textbf{Fs}||||\hat{\textbf{Fs}}||},
\end{aligned}
\end{equation}
and the SSIM $\eta$ can be computed as
\begin{equation}
\begin{aligned}
	    \eta = \frac{(2\mu_{\textbf{Fs}}\mu_{\hat{\textbf{Fs}}}+c_1)(2\sigma_{\textbf{Fs}\hat{\textbf{Fs}}}+c_2)}{(\mu_{\textbf{Fs}}^2+\mu_{\hat{\textbf{Fs}}}^2+c_1)(\sigma_{\textbf{Fs}}^2+\sigma_{\hat{\textbf{Fs}}}^2+c_2)},
\end{aligned}
\end{equation}
where $\mu$ is the mean value, $\sigma$ represents the variance, and $c$ is the default constant to stabilize the division. 
Note that as for the SSIM metric, we compute the SSIM value of real part and imaginary part of spectrum separately and present the average value among those two channels as SSIM only supports the computation of single-channel real vectors.
\subsection{Performance of CRNet in different SNR}

\begin{figure*}
    \centering
    \includegraphics[width=0.90\linewidth]{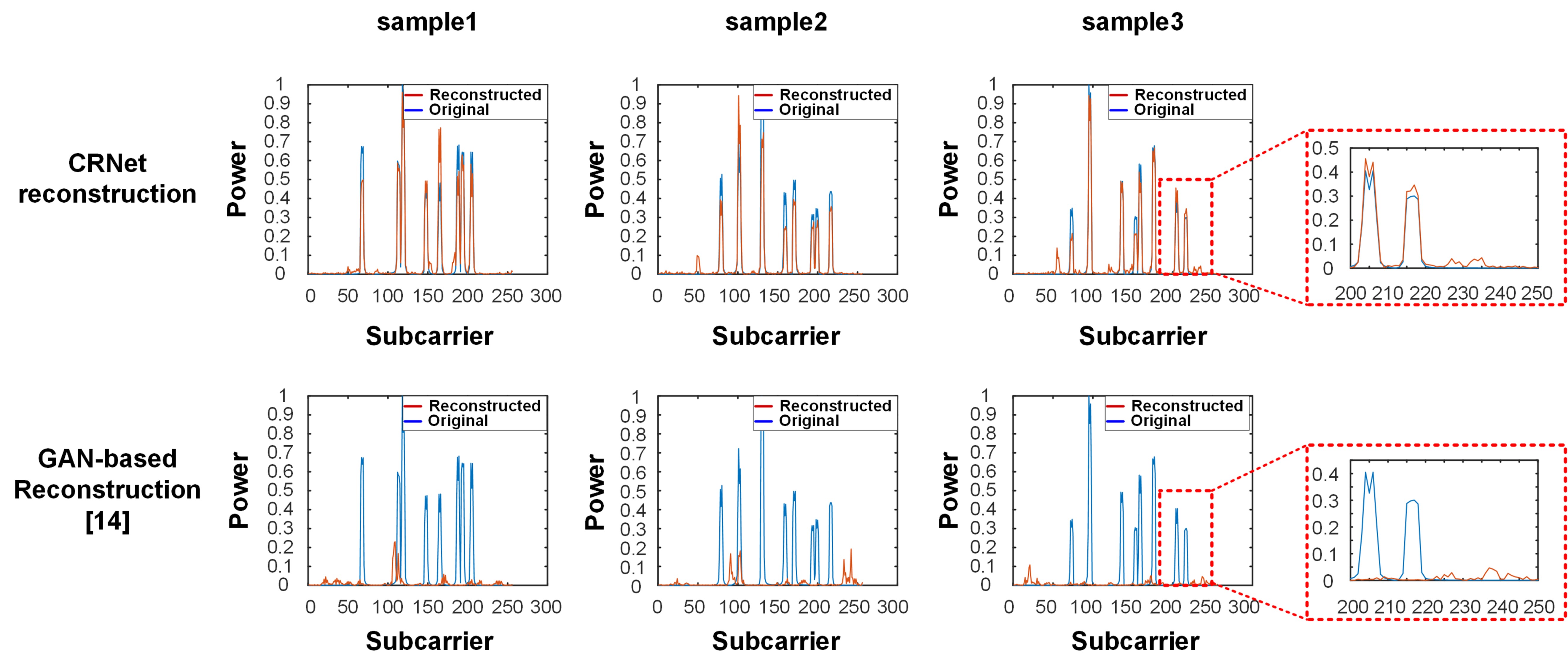}
    \caption{The magnitude reconstruction results of different algorithms on testing samples with SNR = 30 dB and compression rate = 0.125. The first row: the reconstruction result of CRNet algorithm; the second row: the reconstruction result of GAN-based algorithm. As the THz channel effect, the reconstruction task is nontrivial. However, CRNet can still produce reconstructed spectrum, which is very close to the original one in such a low compression rate.}
    \label{fig:Reconstruction}
\end{figure*}
Table I presents the achieved performance of the different algorithms for various signal-to-noise ratios (SNR).
As shown in Table I, although considering THz channel effect increases the difficulty of performing successful spectrum reconstruction, the CRNet significantly outperforms the GAN-based algorithm. The results demonstrate the advantage of the joint optimization between compression and reconstruction from the CRNet towards the DL model architecture design. The gap between the achieved performance of two algorithms becomes more compelling with higher SNR. 

\subsection{Performance of CRNet in different compression rate}
Table II presents the achieved performance of different algorithms in different compression rates. One can notice that the reconstruction quality of GAN-based algorithm degenerates with the smaller compression rate. As an alternative, CRNet shows robustness in different compression rates as the credit of the structured sensing matrix design utilizing fewer under-sampled measurements to perform successful spectrum reconstruction. 

\subsection{Illustration of spectrum reconstruction quality of different algorithms}

Finally, Fig. 3 illustrates some CRNet and GAN-based reconstruction results from  testing samples. Obviously, even in the lower compression rate scenario, CRNet can still offer reconstruction result with high quality due to the joint optimization of compression and reconstruction. On the other hand, the GAN-based algorithm fails to perform valid reconstruction as it struggles to learn the pattern of signal with THz channel effect. 
It is also noteworthy that the number of trainable weights of GAN-based algorithm is 559,587 (the sum of generator and discriminator) and that is 246,499 in the proposed CRNet DL-based algorithm. In other words, the training overhead is reduced by about $56\%$ when the superior performances in the above simulations are reported.

\begin{table}[]
\caption {Comparison of spectrum reconstruction in different SNR with compressed rate = 0.5}

\begin{tabular}{c|cc|cc|cc}
SNR (dB) & \multicolumn{2}{c|}{20}   & \multicolumn{2}{c|}{25}  & \multicolumn{2}{c}{30}  \\ \hline
Method   & GAN    & CRNet              & GAN    & CRNet              & GAN    & CRNet              \\ \hline
MSE      & 0.0329 & \textbf{0.0241} & 0.0304 & \textbf{0.0193} & 0.0325 & \textbf{0.0145} \\ \hline
\begin{tabular}[c]{@{}c@{}}Cosine\\ similarity\end{tabular}      & 0.4659 & \textbf{0.7828} & 0.6588 & \textbf{0.8315} & 0.7511 & \textbf{0.9107} \\ \hline
SSIM     & 0.4745 & \textbf{0.5482} & 0.5184 & \textbf{0.6737} & 0.5090 & \textbf{0.7769}
\end{tabular}
\end{table}

\begin{table}[]
\caption {Comparison of spectrum reconstruction in different compressed rate with SNR = 30dB}

\begin{tabular}{c|cc|cc|cc}
\begin{tabular}[c]{@{}c@{}}Compressed\\ rate\end{tabular} & \multicolumn{2}{c|}{0.5}  & \multicolumn{2}{c|}{0.25} & \multicolumn{2}{c}{0.125} \\ \hline
Method                                                    & GAN    & CRNet              & GAN     & CRNet              & GAN     & CRNet               \\ \hline
MSE                                                       & 0.0325 & \textbf{0.0145} & 0.0347  & \textbf{0.0148} & 0.0357  & \textbf{0.0146}  \\ \hline
\begin{tabular}[c]{@{}c@{}}Cosine\\ similarity\end{tabular}                                                      & 0.7511 & \textbf{0.9107} & 0.3743  & \textbf{0.9346} & 0.2136  & \textbf{0.9161}  \\ \hline
SSIM                                                      & 0.5090 & \textbf{0.7769} & 0.4506  & \textbf{0.7565} & 0.4204  & \textbf{0.7535} 
\end{tabular}
\end{table}

\section{Conclusion}
This paper proposes the CRNet for enhanced spectrum management for THz communications. From the literature review, this work is the first to consider the channel effect, especially THz channel properties, to truly evaluate the achieved performance of different spectral sensing algorithms. CRNet efficiently and effectively performs joint design of compression and reconstruction in an end-to-end learning manner. Simulation results reveal that CRNet outperforms existing algorithms and can provide a realistic reconstruction result as the structured sensing matrix design and corresponding reconstruction module design. To be more specific, CRNet can offer superior performance with only 44$\%$ training overhead as compared to existing DL-based solutions. 
It is noteworthy that CRNet assumes that blind spectrum reconstruction is performed in this paper, which means we can only obtain information from the under-sampled measurements. In our future work, we aim to future consider the case when some of the additional information (e.g., user locations, channel statistics) are provided to further improve the spectrum reconstruction. 



\section*{Acknowledgement}
This work was supported in part by the North Carolina Department of Transportation (NCDOT) under Award TCE$2020$-$03$ and in part by the AC21 Special Project Fund.

\ifCLASSOPTIONcaptionsoff
  \newpage
\fi

\bibliography{references}

\begin{thebibliography}{10}
\providecommand{\url}[1]{#1}
\csname url@samestyle\endcsname
\providecommand{\newblock}{\relax}
\providecommand{\bibinfo}[2]{#2}
\providecommand{\BIBentrySTDinterwordspacing}{\spaceskip=0pt\relax}
\providecommand{\BIBentryALTinterwordstretchfactor}{4}
\providecommand{\BIBentryALTinterwordspacing}{\spaceskip=\fontdimen2\font plus
\BIBentryALTinterwordstretchfactor\fontdimen3\font minus
  \fontdimen4\font\relax}
\providecommand{\BIBforeignlanguage}[2]{{%
\expandafter\ifx\csname l@#1\endcsname\relax
\typeout{** WARNING: IEEEtran.bst: No hyphenation pattern has been}%
\typeout{** loaded for the language `#1'. Using the pattern for}%
\typeout{** the default language instead.}%
\else
\language=\csname l@#1\endcsname
\fi
#2}}
\providecommand{\BIBdecl}{\relax}
\BIBdecl

\bibitem{5g}
A.~Ghosh, A.~Maeder, M.~Baker, and D.~Chandramouli, ``5g evolution: A view on
  5g cellular technology beyond 3gpp release 15,'' \emph{IEEE Access}, vol.~7,
  pp. 127\,639--127\,651, 2019.

\bibitem{Our1}
C.-H. Lin, Y.-T. Lee, W.-H. Chung, S.-C. Lin, and T.-S. Lee, ``Unsupervised
  resnet-inspired beamforming design using deep unfolding technique,'' in
  \emph{IEEE Global Communications Conference (GLOBECOM)}.\hskip 1em plus 0.5em
  minus 0.4em\relax IEEE, 2020.

\bibitem{BF}
W.~Xiong, J.~Lu, X.~Tian, G.~Chen, K.~Pham, and E.~Blasch, ``Cognitive radio
  testbed for digital beamforming of satellite communication,'' in
  \emph{Cognitive Communications for Aerospace Applications Workshop
  (CCAA)}.\hskip 1em plus 0.5em minus 0.4em\relax IEEE, 2017, pp. 1--5.

\bibitem{mmWave2}
S.-C. Lin and I.~F. Akyildiz, ``Dynamic base station formation for solving nlos
  problem in 5g millimeter-wave communication,'' in \emph{IEEE International
  Conference on Computer Communications (INFOCOM)}.\hskip 1em plus 0.5em minus
  0.4em\relax IEEE, 2017, pp. 1--9.

\bibitem{CR}
G.~Wang, K.~Pham, E.~Blasch, T.~M. Nguyen, D.~Shen, X.~Tian, and G.~Chen,
  ``Cognitive radio unified spectral efficiency and energy efficiency trade-off
  analysis,'' in \emph{IEEE Military Communications Conference (MILCOM)}.\hskip
  1em plus 0.5em minus 0.4em\relax IEEE, 2015, pp. 244--249.

\bibitem{Lin4}
S.-C. Lin and H.~Narasimhan, ``Towards software-defined massive mimo for 5g\&b
  spectral-efficient networks,'' in \emph{IEEE International Conference on
  Communications (ICC)}.\hskip 1em plus 0.5em minus 0.4em\relax IEEE, 2018, pp.
  1--6.

\bibitem{THz1}
C.~Han and I.~F. Akyildiz, ``Distance-aware bandwidth-adaptive resource
  allocation for wireless systems in the terahertz band,'' \emph{IEEE
  Transactions on Terahertz Science and Technology}, vol.~6, no.~4, pp.
  541--553, 2016.

\bibitem{Lin1}
S.-C. Lin and K.-C. Chen, ``Cognitive and opportunistic relay for qos
  guarantees in machine-to-machine communications,'' \emph{IEEE Transactions on
  Mobile Computing}, vol.~15, no.~3, pp. 599--609, 2015.

\bibitem{Lin2}
------, ``Improving spectrum efficiency via in-network computations in
  cognitive radio sensor networks,'' \emph{IEEE Transactions on wireless
  communications}, vol.~13, no.~3, pp. 1222--1234, 2014.

\bibitem{CSSS2}
Y.~Fang, L.~Li, Y.~Li, H.~Peng, and Y.~Yang, ``Low energy consumption
  compressed spectrum sensing based on channel energy reconstruction in
  cognitive radio network,'' \emph{Sensors}, vol.~20, no.~5, p. 1264, 2020.

\bibitem{SS4}
J.~Lu, L.~Li, D.~Shen, G.~Chen, B.~Jia, E.~Blasch, and K.~Pham, ``Dynamic
  multi-arm bandit game based multi-agents spectrum sharing strategy design,''
  in \emph{IEEE/AIAA 36th Digital Avionics Systems Conference (DASC)}.\hskip
  1em plus 0.5em minus 0.4em\relax IEEE, 2017, pp. 1--6.

\bibitem{Lin3}
M.~F. Pervej and S.-C. Lin, ``Eco-vehicular edge networks for connected
  transportation: A distributed multi-agent reinforcement learning approach,''
  in \emph{Proc. IEEE 92nd Vehicular Technology Conference (VTC-Fall)}, 2020.

\bibitem{SS3}
G.~Wang, G.~Chen, D.~Shen, X.~Tian, K.~Pham, and E.~Blasch, ``Spread spectrum
  design for aeronautical communication system with radio frequency
  interference,'' in \emph{IEEE/AIAA 34th Digital Avionics Systems Conference
  (DASC)}.\hskip 1em plus 0.5em minus 0.4em\relax IEEE, 2015, pp. 2F1--1.

\bibitem{DLSS1}
X.~Meng, H.~Inaltekin, and B.~Krongold, ``End-to-end deep learning-based
  compressive spectrum sensing in cognitive radio networks,'' in \emph{IEEE
  International Conference on Communications (ICC)}.\hskip 1em plus 0.5em minus
  0.4em\relax IEEE, 2020, pp. 1--6.

\bibitem{Res2}
J.~Ma, S.-C. Lin, H.~Gao, and T.~Qiu, ``Automatic modulation classification
  under non-gaussian noise: A deep residual learning approach,'' in \emph{IEEE
  International Conference on Communications (ICC)}.\hskip 1em plus 0.5em minus
  0.4em\relax IEEE, 2019, pp. 1--6.

\bibitem{config1}
H.~He, C.-K. Wen, S.~Jin, and G.~Y. Li, ``Deep learning-based channel
  estimation for beamspace mmwave massive mimo systems,'' \emph{IEEE Wireless
  Communications Letters}, vol.~7, no.~5, pp. 852--855, 2018.

\end{thebibliography}
\bibliographystyle{IEEEtran}

\end{document}